\documentclass[aps,amsmath,amssymb,prl,twocolumn,showpacs,nofootinbib,flushbottom]{revtex4-1}
\usepackage{epsfig}
\usepackage{graphicx, amssymb}
\usepackage{subfigure}

\usepackage[usenames]{color}

\newcommand{\bk}{\mathbf{k}}

\newcommand{\dg}{\dagger}
\newcommand{\bS}{\mathbf{S}}

\begin{document}

(Accepted by Nature Scientific Reports: \\Jan. 29, 2013)

\title{Phase Diagram for Magnon Condensate in Yttrium Iron Garnet film}

\author{Fuxiang Li$^1$}\author{W.M. Saslow$^1$} \author{V.L. Pokrovsky $^{1,2}$}\affiliation{$^1$Department of Physics, Texas A\&M University, College Station, Texas77843-4242}\affiliation{$^2$Landau Institute for Theoretical Physics, Chernogolovka, Moscow District,142432, Russia}

%\author{Fuxiang Li$^1$, Wayne M. Saslow$^1$$^{,\ast}$, Valery L. Pokrovsky $^{1,2}$\\
%\normalsize{$^{1}$Dept. of Physics, Texas A\&M University, College Station, TX 77843-4242}\\
%\normalsize{$^{2}$Landau Institute for Theoretical Physics, Chernogolovka, Moscow District
%142432, Russia}\\
%\normalsize{$^\ast$To whom correspondence should be addressed:  wsaslow@tamu.edu}	}

%\baselineskip24pt

\date{\today}
%\pacs{75.30.Ds, % spin waves
%03.75.Nt,  %other BEC phenomena
%05.30.Jp, %quantum statistical mechanis Bose systems
%05.30.Pt %Quantum phase transition
%}

\begin{abstract}
{Recently, magnons, which are quasiparticles describing the collective motion of spins, were found to undergo Bose-Einstein condensation (BEC) at room temperature in films of Yttrium Iron Garnet (YIG). Unlike other quasiparticle BEC systems, this system has a spectrum with two degenerate minima, which makes it possible for the system to have two condensates in momentum space. Recent Brillouin Light scattering studies for a microwave-pumped YIG film of thickness $d=5$ $\mu$m and field $H=1$ kOe find a low-contrast interference pattern at the characteristic wavevector $Q$ of the magnon energy minimum. In this report, we show that this modulation pattern can be quantitatively explained as due to non-symmetric but coherent Bose-Einstein condensation of magnons into the two energy minima. Our theory predicts a transition from a high-contrast symmetric phase to a low-contrast non-symmetric phase on varying the $d$ and $H$, and a new type of collective oscillations.  }
\end{abstract}


\maketitle

\newpage
Bose-Einstein condensation (BEC), one of the most intriguing macroscopic quantum phenomena, has been observed in equilibrium systems of Bose atoms, like $^4$He \cite{Kapitza38, Allen38}, $^{87}$Rb \cite{Anderson95} and $^{23}$Na \cite{Davis95}. Recent experiments have extended the concept of BEC to non-equilibrium systems consisting of photons \cite{Kalers10} and of quasiparticles, such as excitons \cite{Butov01}, polaritons \cite{Kasprzak06, Balili07, Amo09}  and magnons \cite{Bunkov08, Demokritov06}. Among these, BEC of magnons in films of Yttrium Iron Garnet (YIG), discovered by the group of Demokritov \cite{Demokritov06,Dzyapko07,Demidov07,Demidov08a, Demidov08b, Dzyapko09, Nowik-Boltyk12}, is distinguished from other quasiparticle BEC systems by its room temperature transition and two-dimensional anisotropic properties. In particular, the spin-wave energy spectrum of a YIG film shows two energetically degenerate minima. Therefore it is possible that the system may have two condensates in momentum space \cite{Leggett01}. An experiment by Nowik-Boltyk {\it et al.} \cite{Nowik-Boltyk12} indeed shows a low-contrast spatial modulation pattern, indicating that there is interference between the two condensates. Current theories \cite{Tupitsyn08, Rezende09, Rezende09b, Rezende10, Malomed10, Troncoso12} do not describe the appearance of coherence or the distribution of the two condensates.

This report points out that a complete description of BEC in microwave-pumped YIG films must account for the $4$th order interactions, including previously neglected magnon-non-conserving terms originating in the dipolar interactions. The theory explains not only the appearance of coherence but also quantitatively explains the  low contrast of the experimentally observed interference pattern.
Moreover, the theory predicts that, on increasing the film thickness $d$ from a small value, there is a transition from a high-contrast symmetric phase S for $d<d_c$, with equal numbers of condensed magnons filling the two minimum states, to a low-contrast coherent non-symmetric phase NS for $d>d_c$, with different numbers of condensed magnons filling the two minimum states. In comparatively thin films ($d<0.2 \mu$m) the same transition can be driven by an external magnetic field $H$.
For $d>d^*$, where $d^*$ is another critical thickness ($d^*>d_c$), the sum of phases of the two condensates changes from $\pi$ to $0$; for $d=d^*$ the system is in a completely non-symmetric phase with only one condensate, for which there is no interference. In the experiment of Ref.\cite{Nowik-Boltyk12} the film thickness $d$ exceeded $d^*$. We suggest that the phase transitions may be identified by measuring the contrast of the spatial interference pattern for various $d$ and $H$. We also predict a new type of collective magnetic oscillation in this system and discuss the possibility of domain walls in non-symmetric phases.

\noindent{\bf Results}

\begin{figure}[htbp]
\centering\includegraphics[width=7cm]{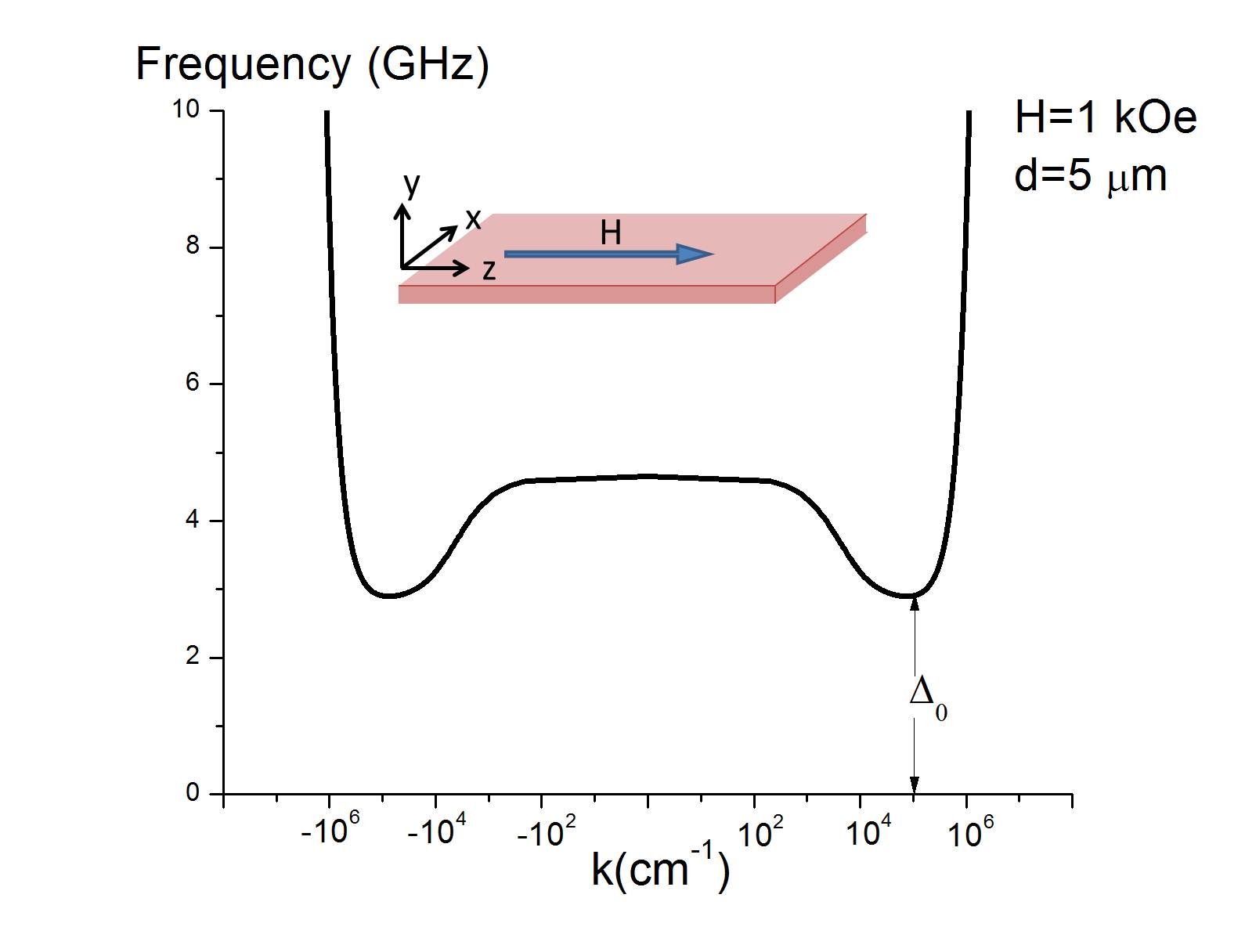}
\caption{ The magnon spectrum in the $k_z$ direction for $d=5$ $\mu$m and $H=1$ kOe. The inset is a schematic diagram of YIG film. }
\label{fig:spectrum}
 \end{figure}

\begin{figure}[htbp]
\centering\includegraphics[width=7cm]{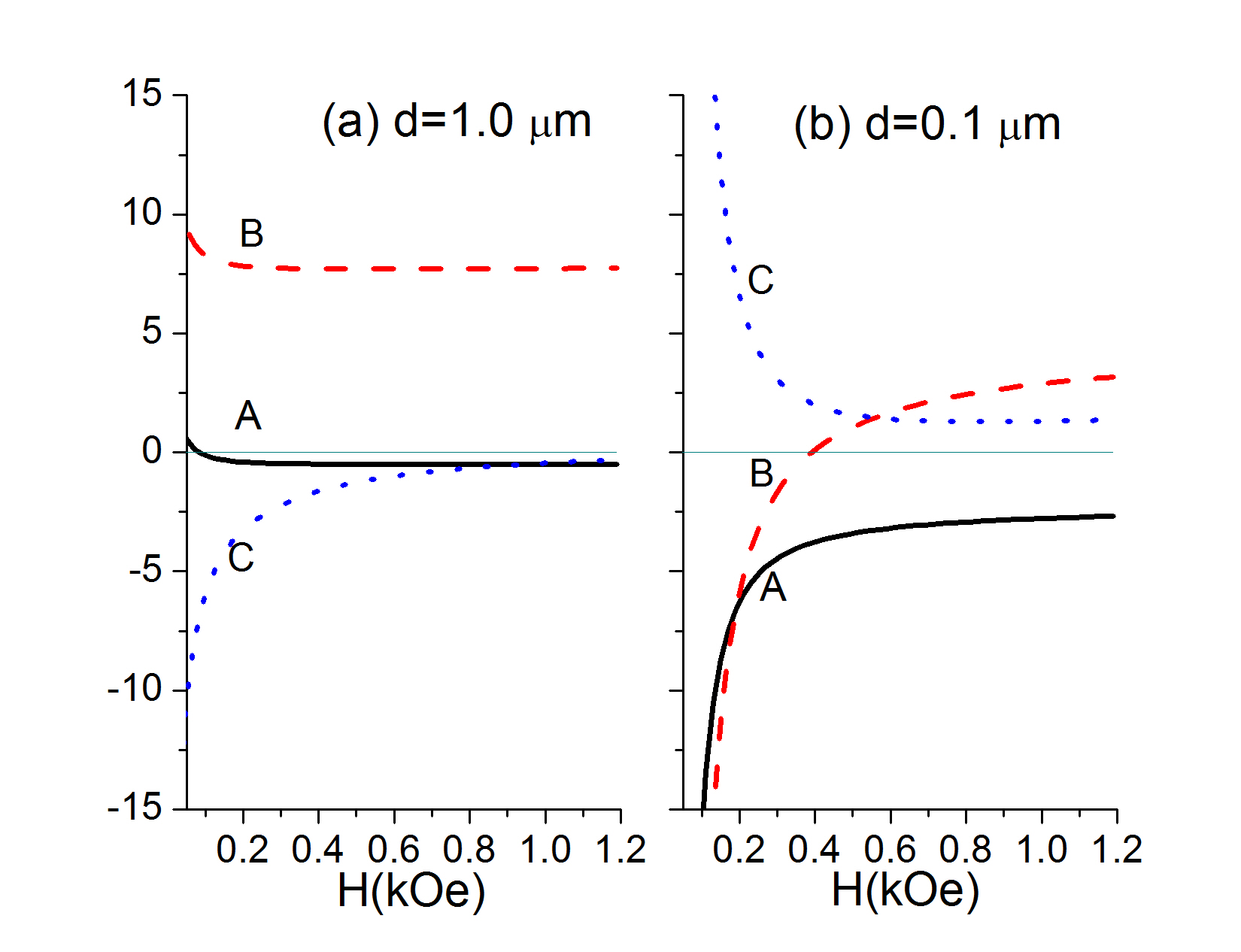}
\caption{Interaction coefficients $A$, $B$ and $C$ (in units of mK/N, with $N$ the total number of spins in the film) as a function of magnetic field $H$ for film thickness (a) $d=1.0$ $\mu$m and (b) $d=0.1$ $\mu$m. }
\label{fig:ABC}
\end{figure}

{\bf Phase Diagram.---}  %
We consider a YIG film of thickness $d$ with in-plane magnetic field $H$ (see inset of Fig. \ref{fig:spectrum}).
The $4$-th order interaction of condensate amplitudes reads \cite{Cherepanov93, Slavin94, Krivosik10}:
 \begin{eqnarray}
 \hat{V}_4&&=A[c_Q^{\dg}c_Q^{\dg}c_Q c_Q + c_{-Q}^{\dg}c_{-Q}^{\dg}c_{-Q}c_{-Q}] \nonumber \\
&&+2B c_{Q}^{\dg}c_{-Q}^{\dg}c_{-Q}c_{Q}  \nonumber \\
&&+C[c_{Q}^{\dg}c_{Q}c_{Q}c_{-Q}+c_{-Q}^{\dg}c_{-Q}c_{-Q}c_{Q}+h.c.]. \label{eqn:interaction}
\end{eqnarray}
Here $c_{\pm Q}$ and $c^{\dag}_{\pm Q}$ are the annihilation and creation operators for magnons in the two condensates located at the two energy minima $(0, \pm Q)$ in the $2$-D momentum space (see. Fig.\ref{fig:spectrum}).
 The coefficients in Eq.(\ref{eqn:interaction}) are:
\begin{eqnarray}
A=&&-\frac{\hbar\omega_M}{4SN}[(\alpha_1-\alpha_3)F_Q-2\alpha_2(1-F_{2Q})] \nonumber \\
&&-\frac{DQ^2}{2SN}[\alpha_1-4\alpha_2],  \nonumber  \\
B=&&\frac{\hbar\omega_M}{2SN}[(\alpha_1-\alpha_2)(1-F_{2Q})-(\alpha_1-\alpha_3)F_Q] \nonumber\\
&&+\frac{DQ^2}{SN}[\alpha_1-2\alpha_2], \nonumber  \\
C=&&\frac{\hbar\omega_M}{8SN}[(3\alpha_1-\frac{20}{3}\alpha_3+3\alpha_2)F_Q \nonumber\\
&&+\frac{16}{3}\alpha_3(1-F_{2Q})]+\frac{DQ^2}{SN}\alpha_3,  \label{eqn:ABC}
\end{eqnarray}
with $\alpha_1=u^4+4u^2 v^2+v^4$, $\alpha_2=2u^2 v^2$ and $\alpha_3=3uv(u^2+v^2)$. Here, $u$ and $v$ are the coefficients of Bogoliubov transformation (see the Methods section for details). $S=14.3$ is the effective spin, $N$ is the total number of spins in the film, $M$ is the magnetization and $\hbar\omega_M=\gamma4\pi M$ with gyromagnetic ratio $\gamma=1.2\times 10^{-5}$eV$/$kOe. $D$ is proportional to the exchange constant and $F_k=(1-e^{-kd})/kd$. Similar results for the coefficients $A$ and $B$ were obtained in Ref.\cite{Tupitsyn08}. The coefficient $C$, which violates magnon number conservation, has not been considered previously. Below we show that $C$ is the only source of coherence between the two condensates. The three coefficients $A$, $B$ and $C$, whose values as functions of $H$ are shown in Fig.\ref{fig:ABC} for two typical values of $d$, determine the distribution of condensed magnons in the two degenerate minima. Ref.\cite{Tupitsyn08}  assumed  a symmetric phase with condensed magnons in both minima having equal amplitudes and equal phases. Later, Ref.\cite{Rezende09} assumed filling of only one minimum. More recently Ref.\cite{Troncoso12} considered Josephson-like oscillations by starting from two condensates with equal numbers of magnons but different phases. Our theory predicts coherent condensates and the ratio of their amplitudes with no additional assumptions.

% The low-energy branches of the spectrum can be considered as two-dimensional.  Due to the competition between the exchange interaction and dipolar interaction, the minimum of the spectrum shifts from $k=0$ to two points $k=\pm Q$.
%In the vicinity of the minimum point, we can approximate the dispersion relation as \cite{Bunkov10}
%\begin{eqnarray}\label{Ep}
%\omega_{\bk}=\Delta_0+\bk \hat{m}^{-1}\bk/2=\Delta_0+\frac{(k_x\pm Q)^2}{m_x}+\frac{k_y^2}{m_y}.
%\end{eqnarray}
 % Here the external magnetic field is along x direction. The gap is $\Delta_0=2.9$ GHz at $H=1kOe$. The position of the minimum is $k_0=5*10^4 cm^{-1}$. The mass of the magnons can be estimated as $m_x\sim k_0^2/\Delta_0$, which is about $6m_e$, and $m_y$ is also of the order of electron mass, but somewhat bigger.

In terms of condensate numbers $N_{\pm Q}$ and phases $\phi_{\pm}$, the condensate amplitudes are $c_{\pm Q}=\sqrt{N_{\pm Q}}e^{i\phi_{\pm}}$.  Substituting them into eq.(\ref{eqn:interaction}) we find:
\begin{eqnarray}
V_4=&&A(N_Q^2+N_{-Q}^2)+2B N_Q N_{-Q} \nonumber \\
&&+ 2C \cos\Phi(N_{Q}^{\frac{3}{2}} N_{-Q}^{\frac{1}{2}}+ N_{Q}^{\frac{1}{2}}N_{-Q}^{\frac{3}{2}}). \label{eqn:V4}
\end{eqnarray}
Clearly the dipole energy depends on the total phase $\Phi=\phi_++\phi_-$. To minimize this energy, for $C>0$ we have $\Phi=\pi$ and for $C<0$ we have $\Phi=0$. Fig.\ref{fig:ABC} shows that the sign of $C$ changes for different $d$ and $H$, which indicates a transition of $\Phi$ between $0$ and $\pi$.  For both $C>0$ and $C<0$ the dipole energy establishes a coherence between the two condensate amplitudes. In contrast to a Josephson-like interaction, the sum rather than the difference of the two condensate phases is fixed.

\begin{figure}[htbp]
\centering\includegraphics[width=7cm]{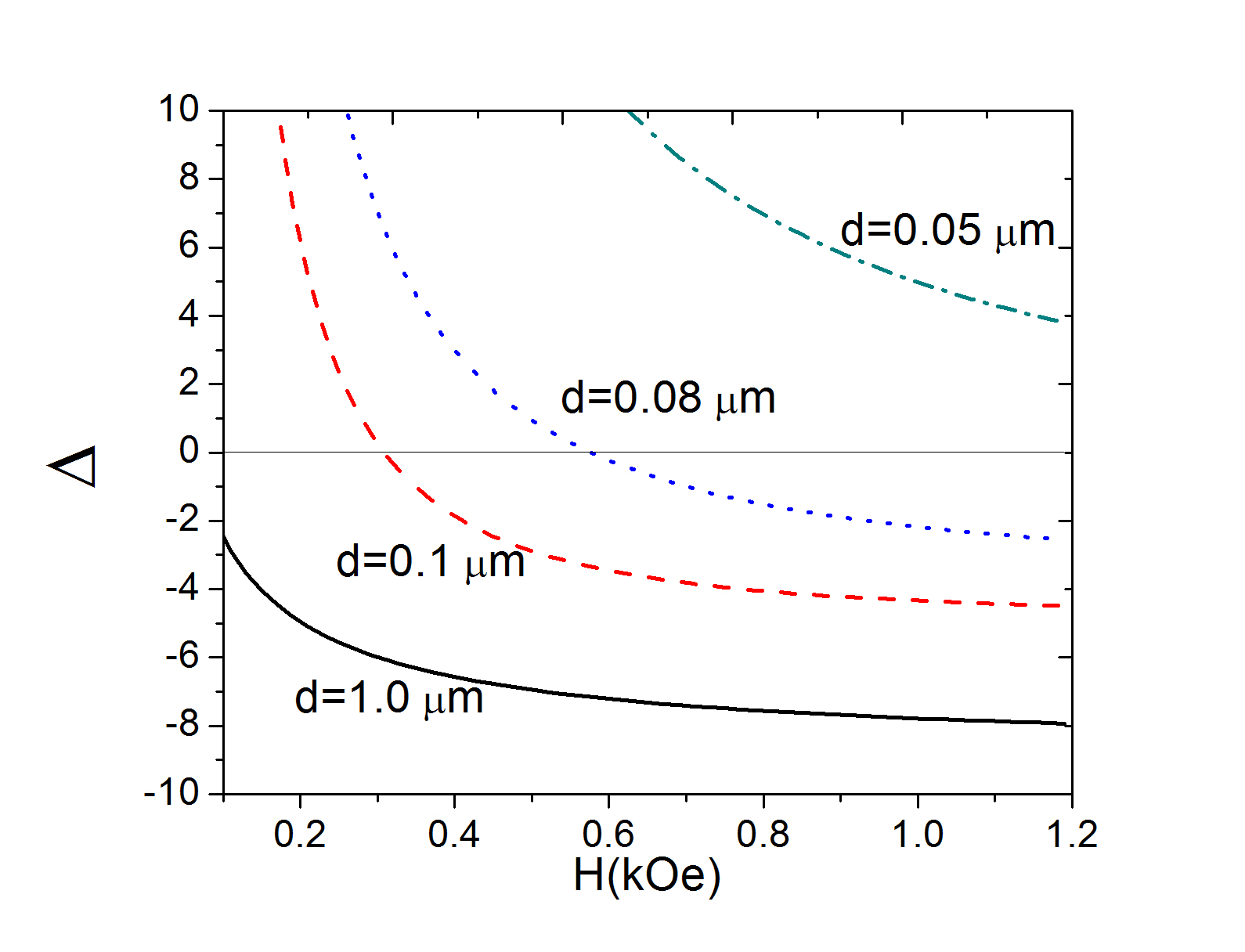}
\caption{Transition criterion $\Delta$ from non-symmetric to symmetric phase, $\Delta$ (in units of mK/N), as a function of magnetic field $H$ for different thicknesses $d$. }
\label{fig:Delta}
\end{figure}

Since the total number of condensed magnons $N_c=N_Q+N_{-Q}$ is uniquely determined by the pumping (see Methods), the energy has only a single free variable, the so far unspecified difference $\delta=N_Q-N_{-Q}$. In terms of $N_c$ and $\delta$ the condensate energy eq.(\ref{eqn:V4}) is:
 \begin{eqnarray}
V_4&&=\frac{1}{2}\Big{[}(A+B)N_c^2-(B-A)\delta^2 \nonumber \\ 
&&-2|C|N_c\sqrt{N_c^2-\delta^2}\Big{]}.
\end{eqnarray}
The ground state of the condensates depends on the criterion parameter $\Delta$, defined as
\begin{eqnarray}
\Delta\equiv A-B+|C|.
\end{eqnarray}
When $\Delta>0$, $\delta=0$ minimizes the energy, so the two minima have equal numbers of condensed magnons. This is the symmetric phase, with $N_Q=N_{-Q}$.  When $\Delta<0$, the minimum shifts to $\frac{\delta^2}{N_c^2}=1-\frac{C^2}{(B-A)^2}$. This is the non-symmetric phase (S), with $N_Q\neq N_{-Q}$. The transition from symmetric to non-symmetric phase (NS) at $\Delta=0$ is of the second order. There is no metastable state of these phases. At $C=0$ one finds $\delta=\pm N_c$, which corresponds to a completely non-symmetric phase with only one condensate. The ground state of the non-symmetric phase is doubly-degenerate, corresponding to the two possible signs for $\delta$.
Fig.\ref{fig:Delta} shows that for a film thickness of about $0.05$ $\mu$m, the symmetric phase is energy favorable up to $H=1.2$ T. For $d=0.08$ $\mu$m, on increasing $H$ to about $0.6$ kOe, there is a transition from symmetric to non-symmetric phase. For the larger thicknesses $d=0.1$ $\mu$m and $d=1$ $\mu$m, the ground state is non-symmetric for $H>0.3$ kOe.  

\begin{figure}[htbp]
\centering\includegraphics[width=7cm]{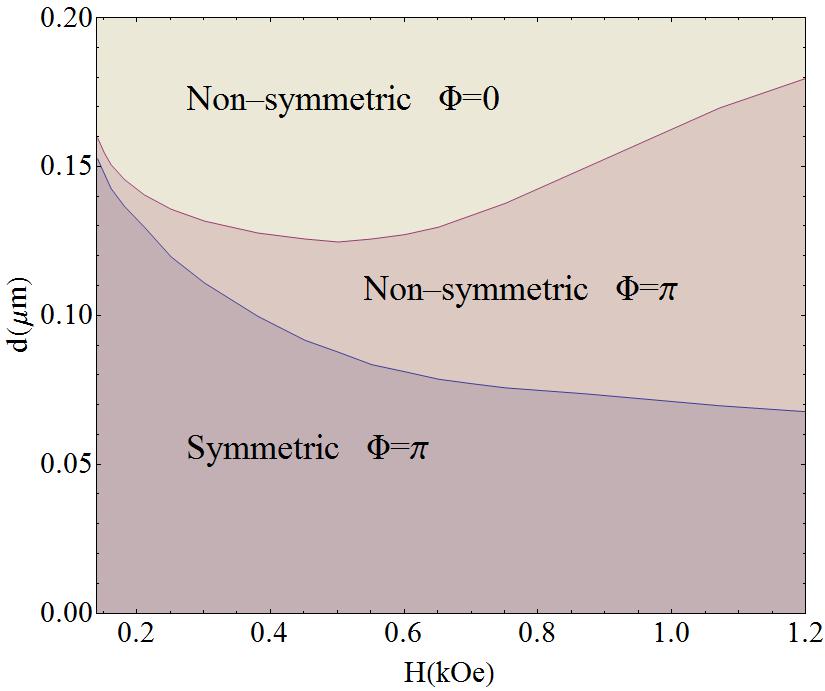}
\caption{The phase diagram for different values of thickness $d$ and magnetic field $H$. }
\label{fig:phase}
\end{figure}

Fig.\ref{fig:phase} gives the phase diagram in $(d, H)$ space. It has three different regions, separated by two critical transition lines, $d_c(H)$ and $d^*(H)$, corresponding to $\Delta=0$ and $C=0$, respectively. For $d=0.13-0.16\,\mu$ the system possesses re-entrant behavior (NS $\Phi=\pi$, to NS $\Phi=0$, to NS $\Phi=\pi$) as $H$ increases.  As shown below, measurement of the contrast, or modulation depth \cite{Nowik-Boltyk12}, of the spatial interference pattern permits identification of the different condensate phases.

\begin{figure}[htbp]
\centering\includegraphics[width=7cm]{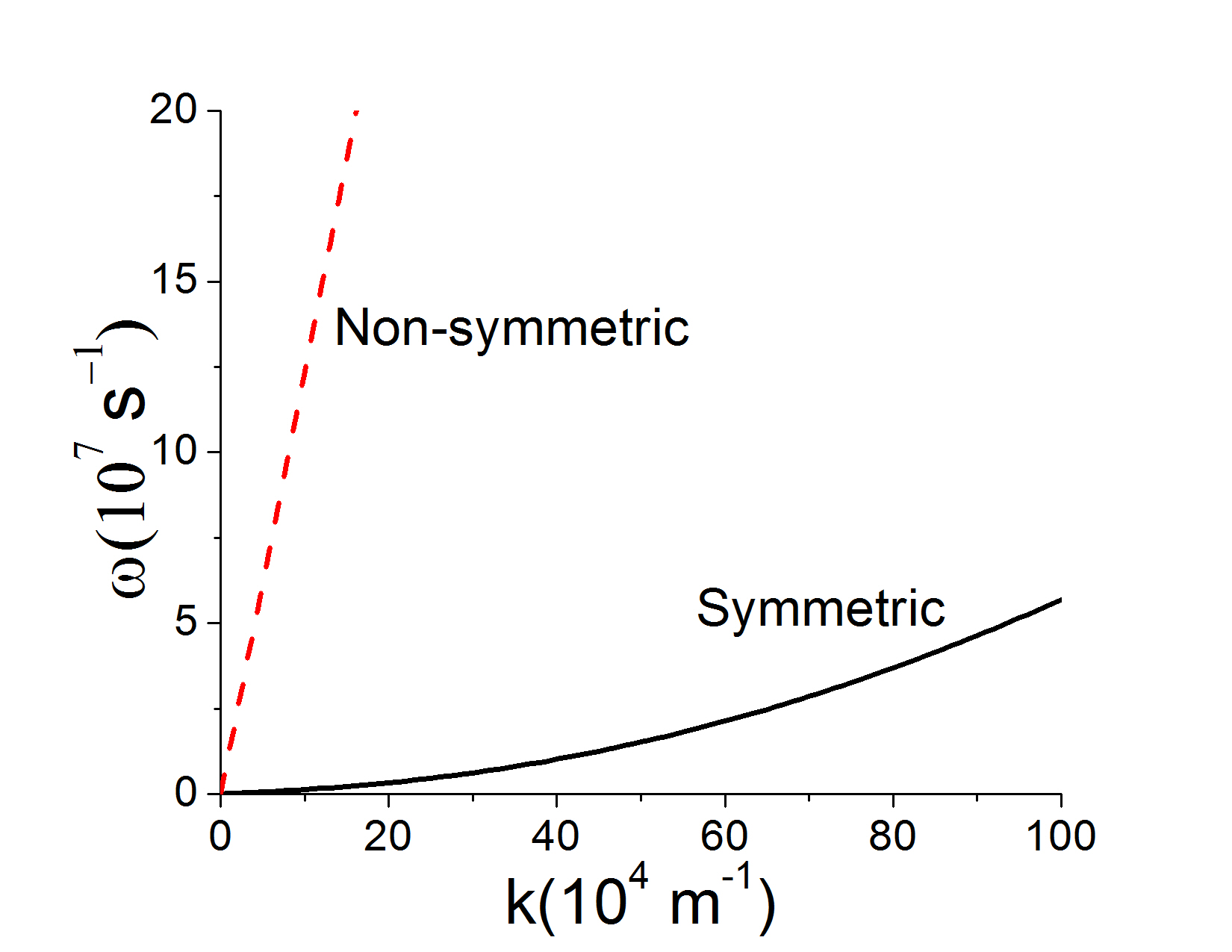}
\caption{ Dispersion of zero sound as a function of wave vector in the direction of external magnetic field for symmetric and non-symmetric cases, respectively. For the non-symmetric case, we choose $H=1$ kOe and $d=5$ $\mu$m. }
\label{fig:ZeroSound}
\end{figure}

{\bf Zero Sound.---}
In two-condensate states the relative phase $\delta\phi = \phi_{+}-\phi_{-}$ is a Goldstone mode. Its oscillation, coupled with the oscillation of the number density $\delta n= n_Q - n_{-Q}$ represents a new type of collective excitation, which we call zero sound (as in Landau's Fermi liquid, this mode is driven by the self-consistent field rather than collisions).
Solving a properly modified Gross-Pitaevskii equation (see Methods), we find its spectrum. In the symmetric phase its dispersion relation is:
\begin{eqnarray}
\omega=\sqrt{\frac{\hbar^2k^4}{4m^2}+N_c\Delta\frac{k^2}{m}}. \label{eqn:zerosound}
\end{eqnarray}
The magnon effective mass is of the order of the electron mass. The density of condensed magnons $n_c=N_c/V$ is about $10^{18}$ cm$^{-3}$ and $\Delta\approx 10$ mK$/$N. The sound speed for small $k$ in this case is $v_{0s}=\sqrt{N_c\Delta /m}$, which is about $100$ m$/$s. Near the transition point $\Delta=0$, the velocity of this zero sound goes to zero.
For the non-symmetric case, the spectrum is:
\begin{eqnarray}
\omega=\sqrt{\frac{\hbar^2k^4}{4m^2}\kappa+N_c(B-A)(\kappa-1)\frac{k^2}{m}},
\end{eqnarray}
where $\kappa\equiv\frac{(B-A)^2}{C^2}$.  In the experiment of Ref.~\cite{Nowik-Boltyk12}, $\kappa\sim 10^{4}$ and $B-A=8.4$ mK$/N$, from which we estimate a sound speed of $3\times 10^{3}$ m$/$s.  The dispersions of zero sound for symmetric and non-symmetric cases are shown in Fig.\ref{fig:ZeroSound}. Note that the range of applicability of the linear approximation decreases significantly for small $C$, where one of the condensate densities is small and the phase fluctuations grow.

{\bf Domain Wall.---}
Since the ground state of the non-symmetric phase is doubly degenerate, it can consist of domains with different signs of $\delta$ separated by domain walls. The width $w$ of such a domain wall is of the order of $\sqrt {\frac{\hbar^2}{2mN_c|\Delta|}}$. For the data of experiment \cite{Nowik-Boltyk12} we estimate that $w\approx 10$ $\mu$m, and a domain wall energy per unit area of about $10^{-9}$J/m$^2$.\\

\noindent{\bf Discussion}

The ground state wave function $\Psi(z)$ generally is a superposition of two condensate amplitudes
$\Psi(z) = (c_Qe^{iQz} + c_{-Q}e^{-iQz})/\sqrt{V}$, where $c_{\pm Q} =\sqrt{N_{\pm Q}}e^{i\phi_{\pm}}$ and $V$ is the volume of the film. 
The spatial structure of $\Psi(z)$ can be measured by Brillouin Light Scattering (BLS), with intensity  proportional to the condensate density $|\Psi|^2=n_Q+n_{-Q}+ 2\sqrt{n_Qn_{-Q}}\cos(2Qz+\phi_+-\phi_-)$.

In their recent experiment, Nowik-Boltyk {\it el al} \cite{Nowik-Boltyk12} observed the interference pattern associated with the ground state. They found that the contrast of this periodic spatial modulation is far below $100\%$; of the order $3\%$. The present theory can quantitatively explain this result. In their experiment,  Ref.\cite{Nowik-Boltyk12} employ $d=5.1$ $\mu$m and $H=1$ kOe. Then eq.(\ref{eqn:ABC}) for $A$, $B$ and $C$ gives  $A=-0.168$ mK$/N$, $B=8.218$ mK$/N$ and $C=-0.203$ mK$/N$, so $\Delta<0$. This corresponds to the non-symmetric phase, where for spontaneous symmetry-breaking with $\delta=N_{Q}-N_{-Q}>0$ the ratio of the numbers of magnons in the two condensates is $\frac{N_{-Q}}{N_{Q}}\approx\frac{C^2}{4(B-A)^2}$. The contrast is $\beta=\frac{|\Psi|^2_{max}-|\Psi|^2_{min}}{|\Psi|^2_{max}+|\Psi|^2_{min}}$. Since $C\ll B$ and $N_{-Q}\ll N_{Q}$, we have $\beta \approx 2\sqrt{\frac{N_{-Q}}{N_{Q}}}\approx \frac{|C|}{|B-A|}$. For the above values of $A$, $B$ and $C$, $\beta$ is of order $2.4\%$, in good agreement with experiment. The smallness of $C$ (and $A$) relative to B is associated with the large parameter $d/l$,  where $l=\sqrt{\frac{D}{\pi \gamma M}}$ is an intrinsic length scale of the system and $l\sim 10^{-6}$ m. In terms of this parameter, $\frac{|C|}{B}\sim (\frac{l}{d})^{2/3}$.

In the experiment of \cite{Nowik-Boltyk12} the contrast $\beta$ reaches a saturation value at a comparatively small pumping power, corresponding to the appearance of BEC. 
This agrees with our expression for $\beta$, which depends only on film thickness $d$ and magnetic field $H$. By varying $d$ and $H$, the contrast can be changed. Specifically, in the symmetric phase, $\beta=1$; in the non-symmetric phase, $\beta<1$; and in the completely non-symmetric case with only one condensate ($d=d^*$), $\beta=0$. Therefore, measurement of the contrast for different values of $d$ and $H$ can give complete information about the phase diagram of the system, for comparison with the present theory.

\begin{figure}[htbp]
\centering\includegraphics[width=7cm]{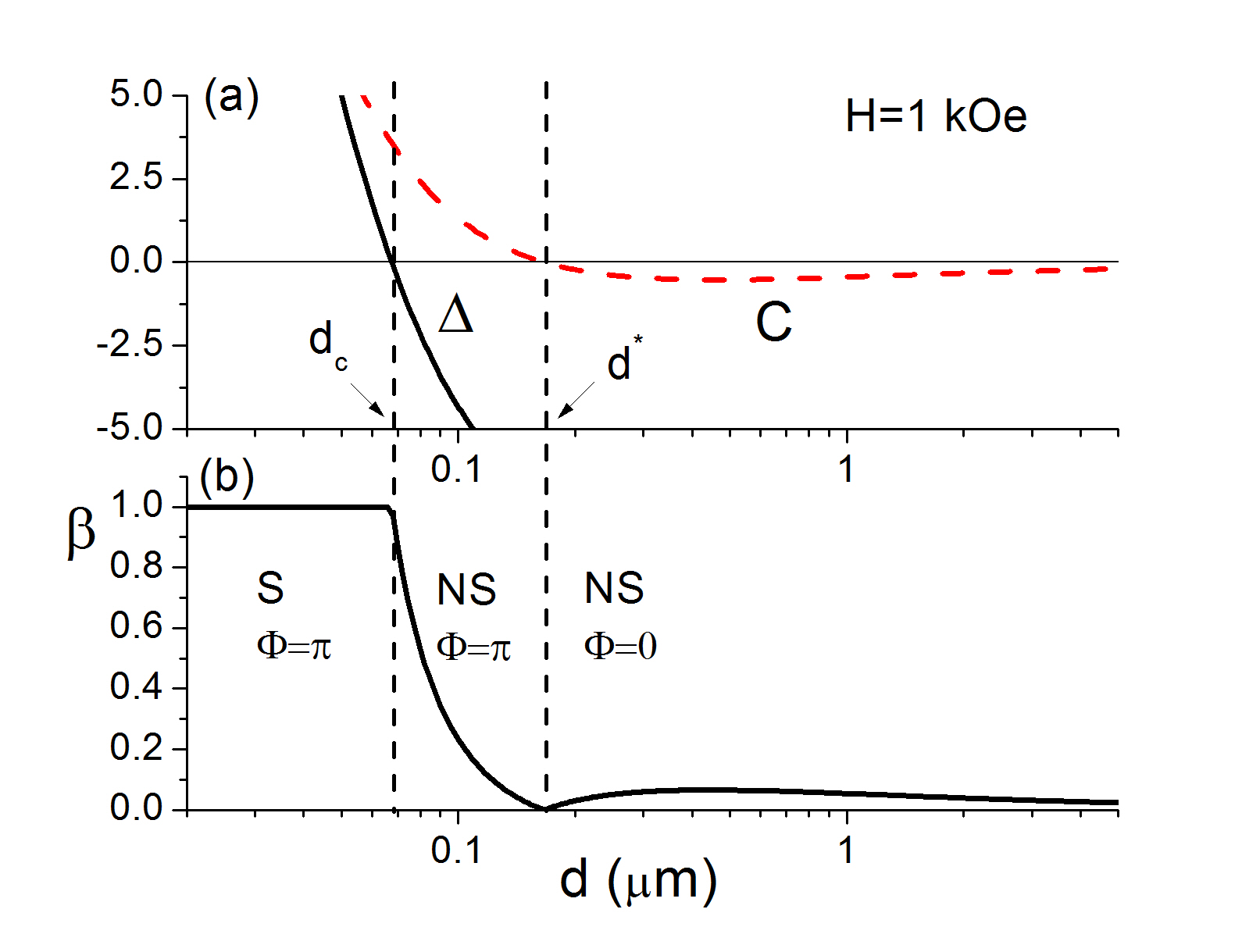}
\caption{(a) Phase transition criterion $\Delta$ and interaction coefficient $C$ as functions of thickness $d$ for fixed magnetic field $H=1$ kOe. (b) The contrast $\beta=\frac{|\Psi|^2_{max}-|\Psi|^2_{min}}{|\Psi|^2_{max}+|\Psi|^2_{min}}$ as a function of thickness $d$ for $H=1$ kOe. S and NS denote symmetric and non-symmetric phase, respectively.  
}
\label{fig:CDB}
\end{figure}

Fig.\ref{fig:CDB} plots $C$, $\Delta$ and $\beta$ as functions of the film thickness $d$ at fixed magnetic field $H=1$ kOe. For small $d$ the system is in the high-contrast symmetric state. On increasing $d$ above $d_c=0.07$ $\mu$m, the sign of $\Delta$ changes, corresponding to the transition from the symmetric to the low-contrast non-symmetric phase. %Note from Fig.\ref{fig:CDB} that the change in contrast occurs in a fairly narrow range of $d$.
As $d$ further increases, to $d^* = 0.17$ $\mu$m, $C$ changes sign, and the total phase $\Phi$ changes from $\pi$ to $0$. Only at this point $d^*$ does the zero-contrast phase (with only one condensate) appear. Correspondingly, a characteristic cusp in the contrast $\beta$ appears near $d^*$.

To conclude, we have calculated the 4-th order
magnon-magnon interactions in the condensate of a film of YIG, including magnon-non-conserving terms  that are responsible for the coherence of two condensates. sFor ufficiently thin YIG films ($d<0.1$ $\mu$) we predict a phase transition from symmetric to non-symmetric phase when the magnetic field exceeds the modest value of $0.2$ kOe.  We also predict
that within the non-symmetric phase there is a thickness $d^*(H)$ where the modulation in the observed interference pattern should totally disappear.\\

\noindent{\bf Methods}

{\bf Magnon Spectrum.---}
In a YIG film with an in-plane external magnetic field $H$, the magnon dispersion has been studied extensively {\cite{Damon61, Sparks70, Kalinikos80}}.
 At low energies, YIG can be described as a Heisenberg ferromagnet with large effective-spin $S=14.3$ \cite{Tupitsyn08, Troncoso12} on a cubic lattice. The Hamiltonian consists of three parts:
 \begin{eqnarray}
{\cal H}=-J\sum_{\langle i,j \rangle} \bS_i\cdot\bS_j+H_D -\gamma H\sum_i S^z_i, \label{eqn:H}
 \end{eqnarray}
 the nearest neighbor exchange energy, the dipolar interaction and the Zeeman energy. We take $y$ to be perpendicular to the film and the magnetic field to be in the plane along $z$.  It is convenient to characterize the exchange interaction by the constant $D=2JSa^2=0.24$ eV${\AA}^2$. The dipolar interaction can be calculated using the method indicated in Refs.\cite{Rezende09, Kalinikos80}. The competition between the dipolar interaction and exchange interaction leads to a magnon spectrum $\omega_k$ with minima located at the two points in 2D wave-vector space given by ${\bf k}=(0,\pm Q)$ (i.e. along $z$), with an energy gap $\Delta_0$. For film thickness $d=5$ $\mu$m and magnetic field $H=1$ kOe, we find that $Q=7.5\times10^{4}$ cm$^{-1}$ and $\Delta_0=2.7$ GHz. In the experiment of \cite{Nowik-Boltyk12} $Q$ was found to be about $3.5\times10^4$ cm$^{-1}$, i.e. about half the predicted value. The reason for this may be associated with a rather shallow energy minimum as a function of wavevector. In such a situation small corrections to our approximate formula can have a large effect on the value of $Q$. The lowest band of the magnon spectrum can be calculated using the Holstein-Primakoff transformation \cite{Madelung}, which expresses the spin operator ${\bf S}$ in terms of boson operators $a$ and $a^\dag$.

To second order in $a$ and $a^\dagger$, the Hamiltonian eq.(\ref{eqn:H}) is:
 \begin{eqnarray}
{\cal H}_0=\sum_{\bk}\Big{[}{\cal A}_ka_k^{\dg}a_k+\frac{1}{2}{\cal B}_k a_k a_{-k}+\frac{1}{2} {\cal B}_k^{*} a_k^{\dg}a_{-k}^{\dg} \Big{]} ,\label{eqn:Hamiltonian}
\end{eqnarray}
with
\begin{eqnarray}
{\cal A}_k&=&\gamma H_0+ Dk^2+\gamma 2\pi M(1-F_k)\sin^2\theta+\gamma 2\pi M F_k   \nonumber \\
{\cal B}_k&=&\gamma 2\pi M(1-F_k)\sin^2\theta-\gamma 2\pi MF_k
\end{eqnarray}
where $F_k \equiv (1-e^{-kd})/kd$ and $M$ is the magnetization of the material ($4\pi M=1.76$ kG). Here, $\theta$ is the angle between the 2D wave vector $\bk$ and the magnetic field direction ($z$).
${\cal H}_0$ of eq.(\ref{eqn:Hamiltonian}) is diagonalized by the Bogoliubov transformation $a_k=u_kc_k+v_kc_{-k}^{\dg}$ with $u_k=(\frac{{\cal A}_k+\hbar\omega_k}{2\hbar\omega_k})^{1/2}$ and $v_k=sgn({\cal B}_k)(\frac{{\cal A}_k-\hbar\omega_k}{2\hbar\omega_k})^{1/2}$, leading to the magnon spectrum:
\begin{eqnarray}
\hbar\omega_k=({\cal A}_k^2-|{\cal B}_k|^2)^{1/2}.
\end{eqnarray}
Fig.\ref{fig:spectrum} gives the magnon spectrum along $k_z$ for typical values of thickness $d$ and magnetic field $H$.

{\bf Number of condensed magnons $N_c=N_Q+N_{-Q}$.---} Experimentally, the spin lattice relaxation time is of order $1$ $\mu$s, whereas the magnon-magnon thermalization time is of order $100$ ns; the magnons are long-lived enough to equilibrate before decaying, thus making BEC possible \cite{Demokritov06}. After the thermalization time the pumped magnons  go to a quasi-equilibrium state with a non-zero chemical potential $\mu$. The number of pumped magnons $N_p=N(T,\mu)-N(T,0)$, where $N(T, \mu)=V\sum_{\bk} \frac{1}{e^{(\omega_k-\mu)/T}-1}$, is determined by the pumping power and the magnon lifetime. $\mu$ is a monotonically increasing function of $N_p$ but cannot exceed the energy gap $\Delta_0$. Therefore, on further pumping $\mu=\Delta_{0}$ and some of the pumped magnons fall into the condensate. The equation $N_{pc}=N(T,\Delta_0)-N(T,0)$ thus defines the critical line for condensation. Since $\Delta_0\ll T$ and $N_p\ll N(T,0)$ this equation can be satisfied at a rather high temperature.
The total number of condensed particles is \cite{Demokritov06, Bunkov10}
\begin{equation}
N_c=N_p-N(T,\Delta_0) +N(T,0)=N(T,\mu)-N(T,\Delta_0).\label{eqn:number}
\end{equation}
In exactly 2D systems BEC formally does not exist since in the continuum approximation the sum in $N(T,\mu)$ diverges. However, for strong enough pumping the chemical potential approaches exponentially close to the energy gap: $\Delta_0-\mu\approx\Delta_0\exp(-N_p/N_0)$, where $N_0= VTm/\hbar^2$. For $N_p/N_0 >\ln (T/\Delta_0)$ all pumped magnons occupy only one or two states $\pm Q$.

Eq.(\ref{eqn:number}) determines only the total number of particles in the condensate. The distribution of the condensate particles between the two minima remains undetermined in the quadratic approximation. To resolve this issue we have shown that it is necessary to consider the fourth order terms in the Holstein-Primakoff expansion of the exchange and dipolar energy. Observe that terms of third order occur in this expansion of the dipolar interaction, but since the total momentum must be zero, such terms vanish for the condensate momenta $(0, \pm Q)$.

{\bf Zero Sound.---} We now provide details about calculating the zero sound spectrum.  We consider small deviations from the static symmetric solution $n_Q=n_{-Q}=n_c/2$, $\phi_+=\pi-\phi_-=0$, so that $n_{\pm Q}=n_{c}/2+\delta n_{\pm}$ with $\delta n_+=-\delta n_-=\delta n/2$ and $\delta \phi_+=-\delta\phi_-=\delta\phi/2$.  Then
\begin{eqnarray}
E=\int dr\Big{(} &&\frac{\hbar^2}{2m}(|\nabla\Psi_+|^2+|\nabla\Psi_-|^2) ) \nonumber \\
&&+ AV(|\Psi_+|^4+|\Psi_-|^4 + 2BV |\Psi_+|^2|\Psi_-|^2 \nonumber \\
&&+ CV (\Psi_+\Psi_-+\Psi_+^*\Psi_-^*)(|\Psi_+|^2+|\Psi_-|^2)\Big{)},\nonumber
\end{eqnarray}
On linearizing, the energy reads:
\begin{eqnarray}
E&&=\int dr \Big{(} \frac{\hbar^2}{4mn_c}|\nabla \delta n|^2+\frac{\hbar^2n_c}{4m}|\nabla\delta \phi|^2 + \frac{\Delta V}{2} \delta n^2 \Big{)}.\nonumber
\end{eqnarray}
Using the commutation relation $[\delta \phi, \delta n]=i$, and the equation of motion  $i\hbar\delta\dot{\phi}=[\delta \phi, H]$, we obtain:
\begin{eqnarray}
\hbar\frac{\partial \delta \phi}{\partial t} &&=-\frac{\hbar^2}{2mn_c}\nabla^2\delta n + \Delta V\delta n, \\
\hbar\frac{\partial \delta n}{\partial t} && =\frac{\hbar^2}{2m}n_c\nabla^2\delta\phi.
\end{eqnarray}
Taking Fourier transforms of the above two equations in coordinate and time, one arrives at the dispersion relations in Eq.(\ref{eqn:zerosound}).

\noindent{\bf Acknowledgements}\\
\noindent{This work was partly supported by the grant of DOE DE-FG02-06ER46278. We are thankful to H. Schulte attracting our attention to the experimental work \cite{Nowik-Boltyk12} and to S. M. Rezende, I.S. Tupitsyn and V. E. Demidov for useful discussion.}\\
\noindent{\bf Author Contributions}\\
\noindent{All authors contributed to the theoretical analysis and the preparation of the manuscript.}\\
\noindent{\bf Additional Information}

There are no competing financial interests.

\end{document}